\documentstyle[aps]{revtex}

\input epsf

\begin{document}

%%%%%%%%%%%%%%%%%%%%%%%%%%%%%%%%%%%%%%%%%%%%%%%%%%%%%%%%%%%%%%%%%%%%%%%%

\draft

\title{Echoing and scaling in Einstein-Yang-Mills critical collapse}

\author{Carsten Gundlach
\thanks{Present address: Max-Planck-Institut f\"ur Gravitationsphysik,
Albert-Einstein-Institut, Schlaatzweg 1, 14473 Potsdam,
Germany. Email: gundlach@aei-potsdam.mpg.de} }

\address{LAEFF-INTA (Laboratorio de Astrof\'\i sica Espacial y F\'\i sica
Fundamental -- Instituto Nacional de Tecnolog\'\i a Aerospacial), 
PO~Box~50727, 28080~Madrid, Spain
}

\date{30 October 1996}

\maketitle

%%%%%%%%%%%%%%%%%%%%%%%%%%%%%%%%%%%%%%%%%%%%%%%%%%%%%%%%%%%%%%%%%%%%%%%%

\begin{abstract}

We confirm recent numerical results of echoing and mass scaling in the
gravitational collapse of a spherical Yang-Mills field by constructing
the critical solution and its perturbations as an eigenvalue
problem. Because the field equations are not scale-invariant, the
Yang-Mills critical solution is asymptotically, rather than exactly,
self-similar, but the methods for dealing with discrete
self-similarity developed for the real scalar field can be
generalized. We find an echoing period $\Delta = 0.73784 \pm 0.00002$
and critical exponent for the black hole mass $\gamma = 0.1964 \pm
0.0007$.

\end{abstract}


%%%%%%%%%%%%%%%%%%%%%%%%%%%%%%%%%%%%%%%%%%%%%%%%%%%%%%%%%%%%%%%%%%%%%%%%

\section{Introduction}

Recently, Choptuik, Chmaj and Bizo\'n \cite{EYM} have studied the
gravitational collapse of an $SU (2)$ Yang-Mills field restricted to
spherical symmetry. Near the boundary in initial data space between
data which form black holes and data which do not (``critical
collapse'') they found two regions with qualitatively different
behavior.  In ``region I'' they found a mass gap, with the minimum
black hole mass equal to the mass of the Bartnik-McKinnon solution,
while in ``region II'' they found the mass scaling and echoing which
are by now familiar from critical collapse in other matter models. The
two kinds of behavior are reminiscent of first and second order phase
transitions.

We review type I and type II behavior in section II. We shall see that
each type of behavior can be understood through the presence of an
intermediate attractor. The type I attractor is the well-known
Bartnik-McKinnon solution, which is static and asymptotically
flat. The type II attractor is self-similar and was not known
before. After having derived the necessary field equations in section
III, we construct it as an eigenvalue problem in section IV of this
paper. In section V we construct its linear perturbations in another
eigenvalue problem and verify that only one of them is growing. This
allows us to calculate the critical exponent governing the mass
scaling semi-analytically, without numerical collapse simulations. In
section VI we summarize our results, which are in good agreement with
collapse simulations, discuss how the Einstein-Yang-Mills system
differs from other systems in which critical collapse has previously
been studied, and put the present paper into perspective.

The analytic and numerical methods of this paper are a generalization
of those developed for the spherical scalar field in
\cite{G1,G2}. In contrast to the scalar
field system the field equations contain a length scale $e^{-1}$ (in
units $c=G=1$), where $e$ is the coupling constant in the
Yang-Mills-covariant derivative $D_a=\nabla_a + eA_a$.  The presence
of a scale in the field equations excludes the existence of an exactly
self-similar solution. Instead we make a series ansatz for a solution
which becomes self-similar asymptotically on spacetime scales much
smaller than $e^{-1}$, or equivalently for curvatures much greater
than $e^2$. The echoing period $\Delta$ is determined by the leading
term of the expansion alone.  For the linearized equations we also
make a series ansatz, but the spectrum $\{\lambda_i\}$ of Lyapunov
exponents is once more determined by the first term of that series
alone. Moreover, to calculate the first term of the perturbation
expansion one only needs to know the first term of the background
expansion. Therefore the higher terms of either expansion are not
required in order to calculate both the echoing period $\Delta$ and
critical exponent $\gamma$ exactly, and will not be calculated here.

%%%%%%%%%%%%%%%%%%%%%%%%%%%%%%%%%%%%%%%%%%%%%%%%%%%%%%%%%%%%%%%%%%%%%%%%

\section{Type I and type II critical phenomena}

Here we summarize the findings of Choptuik, Chmaj and Bizo\'n
\cite{EYM} and explain them in dynamical systems terms. The purpose of
this section is to show that critical phenomena are, in hindsight,
easy to explain, to stress the mathematical similarities between type
I and type II critical behavior, and to motivate the more technical
calculations in the following sections.

For introductory reviews of critical collapse, see
\cite{Bizon,MatGrav,Gund}. Very briefly, one wants to study the limit
in phase space between initial data which eventually form a black hole
and data which do not. Choptuik \cite{Chop} pioneered the method of
(numerically) evolving initial data taken from one-parameter families
of initial data which cross this boundaries, families of data, in
other words, which form a black hole for large values of the
parameter, $p$, but not for small values. Generic families have this
property. By bisection one can numerically determine the critical
value $p_*$ of $p$ for a given family.

For the spherically symmetric massless scalar field Choptuik found
that the black hole mass could be made arbitrarily small, and scaled
like $M\sim (p-p_*)^\gamma$, with $\gamma\simeq 0.37$ the same for all
families of data. Furthermore, before forming the black hole, the time
evolution from all data with $|p-p_*|$ sufficiently small, from all
families, approached one universal solution with the strange property
of being periodic in the logarithm of both $r$ and $t$, or
$\phi(r,t)=\phi(e^\Delta r,e^\Delta t)$, with a period of $\Delta
\simeq 3.44$.  The smaller $|p-p_*|$, the more ``echos'' were visible
before the black hole formed or before the fields dispersed to infinity.

For the spherically symmetric $SU(2)$ Yang-Mills field, Choptuik,
Chmaj and Bizo\'n
\cite{EYM} found the same behavior, with $\Delta\simeq0.74$ and
$\gamma\simeq0.20$ in some region of initial data space, and called it
``type II behavior'' because the black hole mass resembles the order
parameter in a second-order phase transition. In another region of
initial data space they found that as $p$ went through $p_*$, the
black hole mass jumped to a finite value instead of showing power law
behavior.  Before the black hole formed, the solution approached a
regular static solution of that mass, and remained there for the
longer the smaller $|p-p_*|$ was. They named this ``type I behavior'',
in analogy with a first-order phase transition.

Without keeping up the sense of mystery any longer, we now explain
these phenomena in dynamical systems terms \cite{Koike1,Koike2,Bizon}
through the presence of an ``intermediate attractor'', a solution
which has precisely one growing perturbation. We shall introduce a
compact notation which focuses on the essential similarities and
dissimilarities of the two types, while hiding the details, and which
will also be useful later on: by $Z$ we denote the vector of variables
of some first-order form of the field equations, such that, for
example, the complete field equations in spherical symmetry can be
compactly written as $F(Z,Z_{,r},Z_{,t})=0$. The presentation is best
begun with type I behavior.

Type I behavior is dominated by the Bartnik-McKinnon solution
\cite{BM}, which is static, spherically symmetric and asymptotically
flat. It has exactly one unstable perturbation mode
\cite{unstable}, which makes it an intermediate attractor in dynamical
systems terms. Let $Z_*^{\text{(I)}}(r)$ denote this solution. As it
depends only on $r$, its general linear perturbation must be of the
form 
$
\delta Z=\sum_{i=0}^\infty C_i \, e^{\lambda_i t}\ \delta_i Z(r), 
$ where the $C_i$ are free constants. There is exactly one unstable
mode, that is $\Re \lambda_1 > 0$ and $\Re \lambda_i < 0$ for
$i=2,3,\dots$. Furthermore, it is known that the final state arising
from initial data $Z_0(r,\epsilon)\equiv Z_*^{\text{(I)}}(r)+\epsilon
\, \delta_1 Z(r)$ is a black hole for one sign of $\epsilon$, and flat
space with outgoing waves for the other. Let $p$ be the parameter of a
one-parameter-family of initial data, such that for $p>p_*$ a black
hole forms, and for $p<p_*$ the solution disperses. Then for $(p-p_*)$
sufficiently small, the time evolution of data from the family enters
an intermediate asymptotic regime of the form
\begin{equation}
Z(r,t)\simeq Z_*^{\text{(I)}}(r) + {\partial C_1^{\text{(I)}} \over
\partial p}(p_*) \ (p-p_*)
\ e^{\lambda_1^{\text{(I)}} t} \ \delta_1^{\text{(I)}} Z(r),
\end{equation}
where the decaying perturbations ($i\ge2$) can already be neglected,
and where we have approximated $C_1=C_1(p)$ linearly. This solution
leaves the intermediate asymptotic regime to form a black hole or
disperse at a time $T$ when the amplitude of the perturbations, $
\partial C_1 /
\partial p \ (p-p_*) \ e^{\lambda_1 T}$, has reached some small
fiducial value $\epsilon$. This gives a lifetime
\begin{equation}
T \sim -\lambda \ln(p-p_*) + c
\end{equation}
of the meta-stable state, where $c$ depends on the one-parameter
family, but where $\lambda = 1/\lambda_1^{\text{(I)}}$ is
universal. This was in fact observed by Choptuik, Chmaj and Bizon, and
used to estimate $\lambda_1^{\text{(I)}}$, in good agreement with
perturbation theory.

The intermediate attractor in regime II is self-similar instead of
static. It is in fact discretely self-similar, but that is not an
essential detail, and for clarity of comparison with regime I we
pretend in the introduction, and only here, that it is continuously
self-similar. We also disregard the fact that the self-similarity is
only asymptotic on small scales. In suitable coordinates
$Z_*^{\text{(II)}}$ then depends only on $r/t$ instead of only on
$r$. Its general linear perturbation must be of the form $
\delta Z=\sum_{i=0}^\infty C_i \, t^{\lambda_i}\ \delta_i Z(r). 
$ Once more there is exactly one growing mode. (This is required for
the explanation we are about to give to work, and we will demonstrate
it explicitly in section IV.)  The intermediate asymptotic regime for
type II behavior is
\begin{equation}
Z(r,t)\simeq Z_*^{\text{(II)}}(r/t) + {\partial C_1^{\text{(II)}}
\over \partial p}(p_*) \ (p-p_*) \ t^{\lambda_1^{\text{(II)}}} \
\delta_1^{\text{(II)}} Z(r/t).
\end{equation}
One now argues from scale-invariance \cite{Koike1,Koike2,HE2,G2} that
the black-hole mass $M$ is proportional to $T$, and obtains for the
black hole mass
\begin{equation}
\ln M = \gamma \ln(p-p_*) + c,
\end{equation}
where $\gamma = -1/\lambda_1^{\text{(II)}}$ is universal, and $c$ is a
family-dependent constant.  

If the scale-invariance is only asymptotic, as it is for scalar
electrodynamics or Einstein-Yang-Mills, the scaling argument to
calculate the black hole mass goes through unchanged \cite{GunMar}. If
the critical solution is discretely self-similar, as for the scalar
field or the model considered here, with an echoing period of $\Delta$
in the logarithm of the length and time scales, a periodic ``wiggle''
\cite{G2} or ``fine structure'' \cite{HodPiran} is superimposed on the
mass scaling law, which becomes
\begin{equation}
\ln M = \gamma \ln(p-p_*) + c + \Psi[\ln(p-p_*) + c/\gamma],
\end{equation}
where $\Psi$ is a universal periodic function with period $\Delta / (2
\gamma)$. (Note that the same family-dependent constant $c$ appears a second
time in the argument of $\Psi$.) The form of the critical solution and
its perturbations is also more complicated, and will be discussed in
section IV.A and V.A respectively.

%%%%%%%%%%%%%%%%%%%%%%%%%%%%%%%%%%%%%%%%%%%%%%%%%%%%%%%%%%%%%%%%%%%%%%%%%%%%

\section{Field equations and scaling variables} 

In this section we introduce coordinates and field variables for the
spherically symmetric Einstein-Yang-Mills system that are adapted to
type II behavior, where scale-invariance plays a crucial role.  In the
following we consider only type II behavior, and no longer write the
index (II). We adopt the conventions and notation of \cite{EYM}, which
include making both the Yang-Mills field and the coordinates $r$ and
$t$ dimensionless by absorbing suitable factors of $G$, $c$ and $e$
into them.

The spherically symmetric spacetime metric is written as
\begin{equation}
ds^2 \equiv -\alpha^2 \, dt^2 + a^2 \, dr^2 + r^2\left(d\vartheta^2 +
\sin^2
\vartheta \, d\varphi^2\right),
\end{equation}
where $a$ and $\alpha$ depend only on $r$ and $t$,
and the Yang-Mills field strength is given in terms of a single potential
$W(r,t)$ as
\begin{equation}
F \equiv dW \wedge (\tau_1 \, d\vartheta + \tau_2 \, \sin\vartheta \,
d\varphi)
- (1-W^2)\tau_3 d\vartheta \wedge \sin\vartheta \,
d\varphi,
\end{equation}
where $\tau_i$ are the Pauli matrices.  In order to
write the field equations in first order form, we define
\begin{equation}
\Phi \equiv W_{,r}, \qquad \Pi\equiv {a\over \alpha} W_{,t}.
\end{equation}
The complete field equations, reduced to spherical symmetry, are
\begin{eqnarray}
r \Phi_{,t} = && r\left({\alpha\over a} \Pi\right)_{,r}, \\
r \Pi_{,t} = && r\left({\alpha\over a} \Phi\right)_{,r}
+ a \alpha r^{-1} W(1-W^2), \\
\label{aprime} r{a_{,r}\over a} = && {1\over2} (1-a^2) + \Phi^2 + \Pi^2 
+{1\over 2} a^2 r^{-2} (1-W^2)^2, \\
r{\alpha_{,r}\over \alpha} = && {1\over2} (a^2-1) + \Phi^2 + \Pi^2 
-{1\over 2} a^2 r^{-2} (1-W^2)^2, \\
\label{adot} r{a_{,t}\over \alpha} = && 2 \Pi \Phi.
\end{eqnarray}
These equations are the Yang-Mills equation, and three of the four
algebraically independent components of the Einstein equations. The
fourth component is obtained by combining derivatives of the other three,
and is therefore redundant.
       
In order to construct a discretely self-similar solutions, we follow
\cite{G1,G2} in defining new coordinates
\begin{equation}
\tau \equiv \ln (-t), \qquad \zeta \equiv \ln\left(-{r\over t}\right) -
\xi_0(\tau), 
\end{equation}	
where $\xi_0$ is a periodic function to be determined, with period
$\Delta$. (This definition differs slightly from \cite{G2} in that $t$
and $r$ are dimensionless, and that $t$ is negative.) The resulting
spacetime metric is
\begin{equation}
ds^2 = e^{2\tau} \left\{ -\alpha^2 \, d\tau^2 + e^{2(\zeta+\xi_0)} 
\left[a^2\left(d\zeta + (1+\xi_0')\,d\tau\right)^2 +    
d\vartheta^2 + \sin^2
\vartheta \, d\varphi^2\right] \right\},
\end{equation}
where $a$ and $\alpha$ are now functions of $\zeta$ and $\tau$, and
where $\xi_0 \equiv \xi_0(\tau)$ and $\xi_0'\equiv d\xi_0/d\tau$.  As
discussed in \cite{G2}, discrete self-similarity is equivalent to $a$
and $\alpha$ being periodic in $\tau$. In the field equations we make
the replacements
\begin{equation}
r{\partial \over \partial r} = {\partial \over \partial \zeta}, \qquad
r{\partial \over \partial t} = e^{\zeta+\xi_0(\tau)} \left[ -{\partial
\over \partial \tau} + (1+\xi_0'(\tau)) {\partial \over \partial
\zeta}\right], \qquad r=e^{\tau + \zeta + \xi_0(\tau)}
\end{equation}
to transform to the new coordinates.

We shall be looking for a solution in which $a$ and $\alpha$ are
periodic. What does this mean for the matter variables $\Phi$, $\Pi$
and $W$? The Einstein equations suggest that $\Phi$ and $\Pi$ should
be periodic too, but $W$ cannot be periodic because of the explicit
presence of the factors $e^\tau$ in the equations.  Nor can we simply
absorb such a factor into the definition of $W$ to make it
periodic. This means that the equations have no nontrivial
self-similar (periodic) solutions. The physical reason is the presence
of the length scale $e^{-1}$ in the problem, which is only hidden by
the dimensionless variables.  Following a suggestion by Matt Choptuik
\cite{Chopprivcom}, we define a new scalar field $S$ by
\begin{equation}
\label{Sdef}
W \equiv 1 - rS.
\end{equation}
With this
definition, the two potential terms arising in the field equations,
\begin{equation}
r^{-1}W(1-W^2) = (1-rS)(2-rS)S, \qquad r^{-2}(1-W^2)^2 = (2-rS)^2 S^2,
\end{equation}
split into the sum of a term which no longer contains $r$ explicitly,
plus terms containing positive powers of $r$, which become negligible
on small spacetime scales (as $r\to 0$ or $\tau \to - \infty$).  Two
further definitions, namely $\Pi_\pm
\equiv \Pi \pm \Phi$ and $g \equiv {a/
\alpha}$, will be useful because $g$ alone determines the ingoing and
outgoing null geodesics, and $\Pi_+$ and $\Pi_-$ are the components of the
matter field propagating along them.

In the following, we use the coordinates $\zeta$ and $\tau$, and the
fields $Z\equiv\{ a, g, \Pi_+, \Pi_-, S\}$. In these variables, the
complete field equations, including the definitions of $\Pi_+$ and
$\Pi_-$ in terms of $S$, are
\begin{eqnarray}
\label{1} \Pi_{\pm,\zeta} = && {\mp e^{\zeta+\xi_0} g \Pi_{\pm,\tau}
+C \Pi_\pm \mp a^2
(1-e^{\tau+\zeta+\xi_0}S) (2-e^{\tau+\zeta+\xi_0}S) S
\over 1 \mp (1+\xi_0') e^{\zeta+\xi_0} g },
\\
\label{5}
a_{,\zeta} = && {a\over 2}\left(C + \Pi_+^2 + \Pi_-^2\right), \\
\label{2} g_{,\zeta} = && C g, \\
\label{6} S_{,\zeta} = && - S - {1\over 2} \left(\Pi_+ - \Pi_-\right), \\
\label{3} 0 = && a_{,\tau} 
+ e^{-(\zeta+\xi_0)} g^{-1} {a\over 2} \left(\Pi_+^2 -
\Pi_-^2\right) 
- (1+\xi_0') {a\over 2}\left(C + \Pi_+^2 + \Pi_-^2\right), \\
\label{4} 0 = && S_{,\tau} 
- e^{-(\zeta+\xi_0)} g^{-1} {1\over 2} \left(\Pi_+ + \Pi_-\right)
+ (1+\xi_0') \left[S + {1\over 2} \left(\Pi_+ - \Pi_-\right)\right],
\end{eqnarray}
where
\begin{equation}
\nonumber
C \equiv 1- a^2 + a^2 (2-e^{\tau+\zeta+\xi_0}S)^2 S^2.
\end{equation}
As suggested by the way we have written the equations, the first five
can be treated as evolution equations in $\zeta$, with periodic
boundary conditions in $\tau$, and the last two as constraints which
are propagated by the evolution equations. Note that now only positive
powers of $e^\tau$ appear explicitly, so that in the limit
$\tau\to-\infty$ we are left with a set of nontrivial, scale-invariant
equations for $Z$. The terms multiplied by $e^\tau$ are ``irrelevant''
in the language of renormalisation group theory \cite{BriKu}.

The equations are invariant under $W\to-W$, and the potential for $W$
has the two minima $W=\pm1$. In (\ref{Sdef}) we have assumed that $W
\to 1$ asymptotically. A solution tending to $W=-1$ can be trivially
obtained from one tending to $W=1$ by changing the sign of $W$, $\Phi$
and $\Pi$, while leaving $S$, $a$ and $\alpha$ unchanged. The field
equations are left unchanged.

%%%%%%%%%%%%%%%%%%%%%%%%%%%%%%%%%%%%%%%%%%%%%%%%%%%%%%%%%%%%%%%%%%%%%%%%

\section{Background solution}

\subsection{The eigenvalue problem}

In this section we construct the solution $Z^{\text{(II)}}_*(r,t)$
which dominates type II behavior. To find a solution which is
asymptotically self-similar in the limit $\tau \to - \infty$, that is
on spacetime scales smaller than the intrinsic scale of the field
equations, we make the ansatz
\begin{equation}
\label{scaling}
Z_*(\zeta,\tau) = \sum_{n=0}^\infty e^{n\tau}\, Z_{*n}(\zeta,\tau),
\end{equation}
where each $Z_{*n}$ is periodic in $\tau$. $Z_{*0}$ is the solution of
a nonlinear eigenvalue problem, with eigenvalue $\Delta$, and boundary
conditions arising from certain regularity requirements. $Z_{*1}$ is
the solution of an inhomogeneous nonlinear boundary value problem,
with source terms depending on $Z_{*0}$.  Similar boundary value
problems completely determine all higher $Z_{*n}$ recursively.

In the following we are interested only in the equations for $Z_{*0}$,
and from now on we suppress the suffix ${}_{*0}$ on the components of
$Z_{*0}$, denoting $\Pi_{+*0}$ simply by $\Pi_+$ etc. (In the compact
formal notation $Z_{*0}$ we keep the suffix.) The equations for
$Z_{*0}$ are derived from those for $Z$ above by setting the factor
$e^\tau$ equal to zero at each explicit occurrence.  We choose to
evolve only $\Pi_+$, $\Pi_-$ and $g$ in $\zeta$, with eqns. (\ref{1})
and (\ref{2}), and to determine $a$ and $S$ at each new value of
$\zeta$ from the constraints, eqns.  (\ref{3}) and (\ref{4}). The
final set of equations, those we have solved numerically, is, in the
simplified notation,
\begin{eqnarray}
\label{background_Pi}
\Pi_{\pm,\zeta} = && {\mp e^{\zeta+\xi_0} g \Pi_{\pm,\tau}
+\left(1 - a^2 + 4 a^2 S^2\right) \Pi_\pm \mp 2 a^2 S
\over 1 \mp (1+\xi_0') e^{\zeta+\xi_0} g },
\\
\label{background_g}
g_{,\zeta} = && \left(1 - a^2 + 4 a^2 S^2\right) g, \\
\label{background_a}
0 = && a_{,\tau} 
+ e^{-(\zeta+\xi_0)} g^{-1} {a\over 2} \left(\Pi_+^2 -
\Pi_-^2\right) 
- (1+\xi_0') {a\over 2}\left(1 - a^2 + 4 a^2 S^2 + \Pi_+^2 + \Pi_-^2\right),
\\
\label{background_S}
0 = && S_{,\tau} - e^{-(\zeta+\xi_0)} g^{-1} {1\over 2} \left(\Pi_+ +
\Pi_-\right) + (1+\xi_0') \left[S + {1\over 2} \left(\Pi_+ -
\Pi_-\right)\right].
\end{eqnarray}
All fields are periodic in $\tau$ with a period $\Delta$ that is to be
determined as an eigenvalue.  Here as in the example of the scalar
field \cite{G2}, the field equations are complemented by regularity
conditions at the center $r=0$ (for $t<0$), and at the past
self-similarity horizon (the past light cone of the point $(r=0,t=0)$,
or $r\simeq -t$).  One can solve these boundary conditions in terms of
free parameters.

To make $r=0 \Leftrightarrow \zeta=-\infty$ a regular center, we
impose $a=1$ and $g=1$ there.  We expand in powers of $e^\zeta$, and
notice that $a$, $g$ and $\Pi$ are even in that expansion (because
they are even in $r$ at $r=0$), while $S$ and $\Phi$ are odd. We label
the orders of this expansion by a suffix in round brackets to
distinguish them from the orders in the expansion (\ref{scaling}).
The expansion coefficients can be given recursively in terms of one
free periodic function $S_{(1)}(\tau)$. To order $e^{3\zeta}$ they
are, giving $\Pi$ and $\Phi$ instead of $\Pi_+$ and $\Pi_-$,
\begin{eqnarray}
\label{left_first}
a_{(0)}(\tau) = && 1, \\
g_{(0)}(\tau) = && 1, \\
\Pi_{(0)}(\tau) = && 0, \\ 
S_{(1)}(\tau) = && {\rm free}, \\ 
\Phi_{(1)}(\tau) = && - 2 S_{(1)}, \\
a_{(2)}(\tau) = && 2 S_{(1)}^2, \\
g_{(2)}(\tau) = && 0, \\
\Pi_{(2)}(\tau) = && e^{\xi_0} \left[S_{(1)}' 
- (1+\xi_0') S_{(1)} \right], \\
S_{(3)}(\tau) = && {1\over 10} \left\{ e^{\xi_0} \left[\Pi_{(2)}' 
- 2 (1+\xi_0') \Pi_{(2)}
\right] + 8 S_{(1)}^3 \right\}, \\
\label{left_last}
\Phi_{(3)}(\tau) = && - 4 S_{(3)}.
\end{eqnarray}
These expressions are used to impose the asymptotic boundary condition at
$\zeta \to -\infty$ at some small value of $\zeta$,
say $\zeta=\zeta_{\text{left}}$. 

We use the remaining coordinate freedom, the choice of $\xi_0(\tau)$, to
move the self-similarity horizon to the coordinate surface $\zeta=0$
through the coordinate condition
\begin{equation}
\label{coordcond}
\left[1 - (1+\xi_0') e^{\xi_0} g\right]_{\zeta=0} = 0,
\end{equation}
which means that $\zeta=0$ is null, and impose analyticity there by the
condition
\begin{equation}
\label{regcond}
\left[ - e^{\xi_0} g \Pi_{+,\tau}
+(1-a^2 + 4a^2 S^2) \Pi_+ - 2 a^2 S\right]_{\zeta=0} = 0.
\end{equation}
(This is a regular and sufficient condition, by the same argument already
used in \cite{G2}.)

These two constraints can be solved recursively after expanding, this time
in powers of $\zeta$. We denote the components of this expansion also by
subscripts in round brackets.  The two free parameters here are the
periodic functions $g_{(0)}(\tau)$
and $\Pi_{-(0)}(\tau)$. From (\ref{coordcond}), one obtains the algebraic
identity
\begin{equation}
\label{g0}
g_{(0)}=[e^{\xi_0}
(1+\xi_0')]^{-1}. 
\end{equation}
To obtain the leading order coefficients of the other fields, we substitute
(\ref{g0}) into eqns. (\ref{background_S}), (\ref{background_a}), and
(\ref{background_Pi}) (upper sign), obtain
\begin{eqnarray}
&& S_{(0)}' + (1+\xi_0') S_{(0)} - (1+\xi_0') \Pi_{-(0)} = 0, \\
&& a_{(0)}' - (1+\xi_0')\left(\Pi_{-(0)}^2 + {1\over2}\right)a_{(0)} 
- (1+\xi_0')\left(2S_{(0)}^2 - {1\over2}\right) a_{(0)}^3 = 0, \\
&& \Pi_{+(0)}' - (1+\xi_0')\left(1-a_{(0)}^2 + 4 a_{(0)}^2 S_{(0)}^2\right)
\Pi_{+(0)}
+ (1+\xi_0') 2
a_{(0)}^2 S_{(0)} = 0,
\end{eqnarray}
and consider these as linear ODEs for 
$S_{(0)}$, $(a_{(0)})^{-2}$, and $\Pi_{+(0)}$ respectively.

As in the scalar field case, we make the assumption that the metric
variables $a$ and $g$ contain only even frequencies in $\tau$, and the
matter variables $\Pi_+$, $\Pi_-$ and $S$ only odd frequencies. This
is compatible with the equations for $Z_{*0}$, but not with the
equations for the general $Z$.  If this symmetry did not hold, the
right-hand side of eqn. (\ref{4}) would contain even terms in $\tau$,
and among them generically a term constant in $\tau$. Then $S$ would
not be periodic in $\tau$, but would have a term linear in $\tau$, and
through the Einstein equations this would be in contradiction to the
periodicity of $\alpha$ and $g$, and hence the self-similarity of
$Z_{*0}$.

The equivalent of the field $S$ here is the scalar field $\phi$ in the
scalar field model, and for a massive or self-interacting $\phi$ a
similar argument holds. The equations for a massless $\phi$, however,
do not contain $\phi$ itself but only its derivatives
$\Pi_\pm$. Therefore a linear dependence of $\phi$ on $\tau$ would not
clash with spacetime self-similarity.  Such solutions exist, and have
been investigated in \cite{Brady}, but surprisingly the critical
solution for the massless field is not of this kind, and the massless
and massive (or self-interacting) scalar field are therefore in the
same universality class.

\subsection{Numerical construction}

Our numerical method has been described in detail elsewhere
\cite{G2}. By decomposing all fields in Fourier components with
respect to $\tau$, the PDEs in $\tau$ and $\zeta$ go over into a
(large) system of ODEs in the variable $\zeta$ for the Fourier
components.  ODEs in $\tau$ alone, in the boundary conditions and the
constraints, go over into algebraic equations which can be solved in
closed form. $\Delta$ now appears as a parameter in the Fourier
transformation of the $\tau$-derivatives.

A solution of the field equations and boundary conditions exists only
for isolated values of $\Delta$, and we have found precisely one. The
convergence radius of our relaxation algorithm is smaller than for the
scalar field, probably because of the shorter period $\Delta$, and
instead of an ad-hoc initial guess we had to use collapse data kindly
provided by Matt Choptuik \cite{Chopprivcom} to obtain a good enough starting
value for the relaxation algorithm.

We find good agreement of $Z_{*0}$ with the $Z$ of a critical collapse
simulation for $-3.00<\tau<-2.22$ \cite{Chopprivcom}, which is not
very surprising as we started our numerical search with these data,
but nevertheless confirms that the ansatz (\ref{scaling}) for $Z_*$ is
consistent and converges for small enough $\tau$, with $Z_{*0}$ the
dominant term.

To obtain error bars on the solution, we have checked convergence with the
numerical parameters, $\zeta_{\text{left}}$, the number $N$ of Fourier
components and the grid spacing $\Delta\zeta$, by varying one of them
at a time. 

\begin{figure}
\centerline{\epsffile{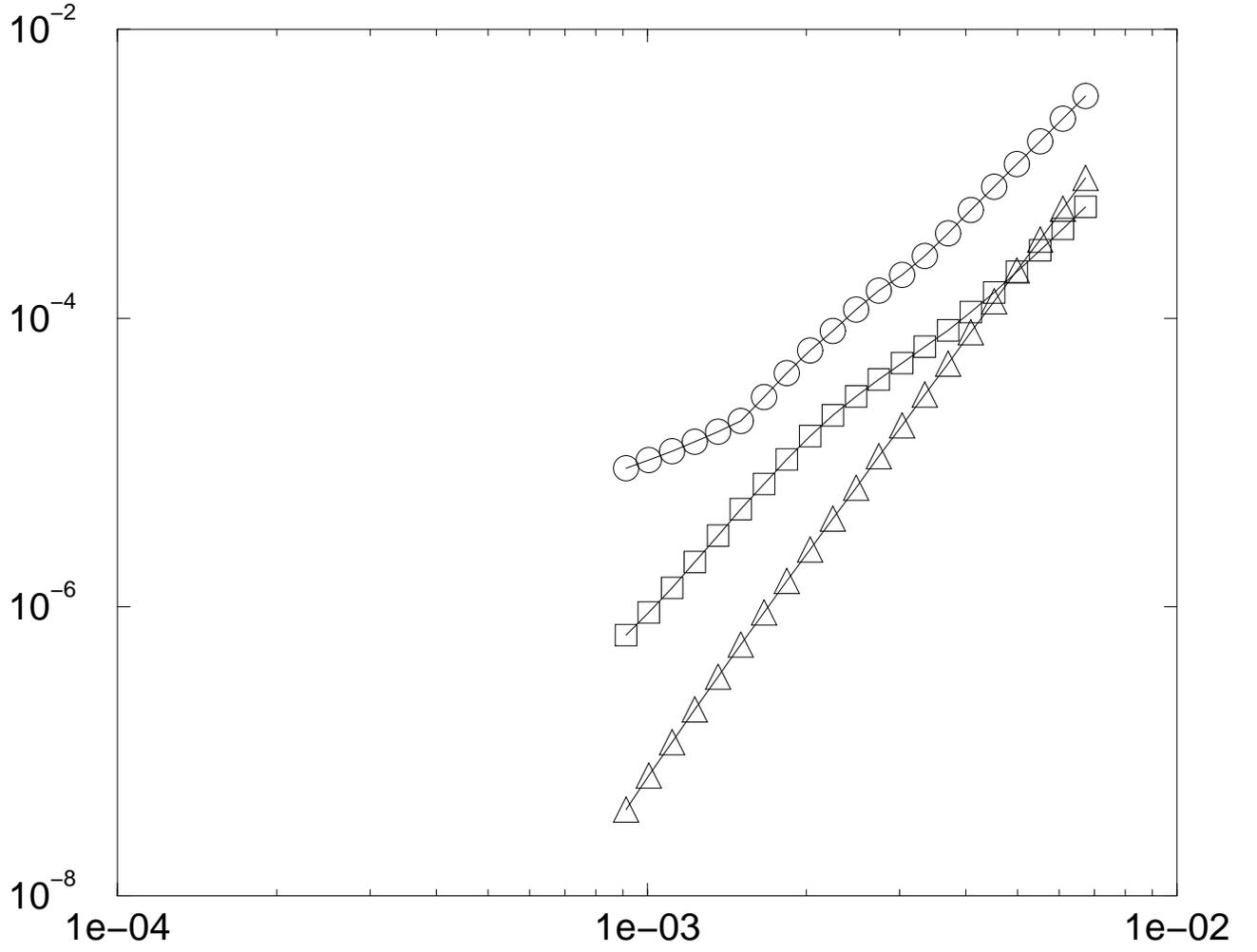}}
\caption{Quartic convergence of  $Z_{*0}$ with
$\exp\zeta_{\text{left}}$. Assuming that the error is $E\simeq A \exp
4\,\zeta_{\text{left}}$, the difference between two numerical solutions
obtained with $\zeta_{\text{left}}$ and $\zeta_{\text{left}} + \Delta
\zeta_{\text{left}}$ is $E'\simeq 4 A \,\Delta\zeta_{\text{left}}\,\exp 4
\,\zeta_{\text{left}}$. Therefore we plot here
$E'/(4\Delta\zeta_{\text{left}}) \simeq E$, against
$\exp\zeta_{\text{left}}$.  Circles denote the maximal error, over all
gridpoints and Fourier components, squares the root mean square error,
and triangles the error in $\Delta$. $N=64$ and $\Delta\zeta=0.1$. The
production value $\zeta_{\text{left}}=-6.4$ corresponds to the fifth
point from the left.}
\label{convergence_zetaleft}
\end{figure}

Fig. \ref{convergence_zetaleft} demonstrates quartic convergence with
$\exp\zeta_{\text{left}}$, as expected from our expansion to order
$O(\exp3\zeta_{\text{left}})$. This convergence breaks down at very
small values of $\exp\zeta_{\text{left}}$, due to the fact that all
fields become very small.

\begin{figure}
\centerline{\epsffile{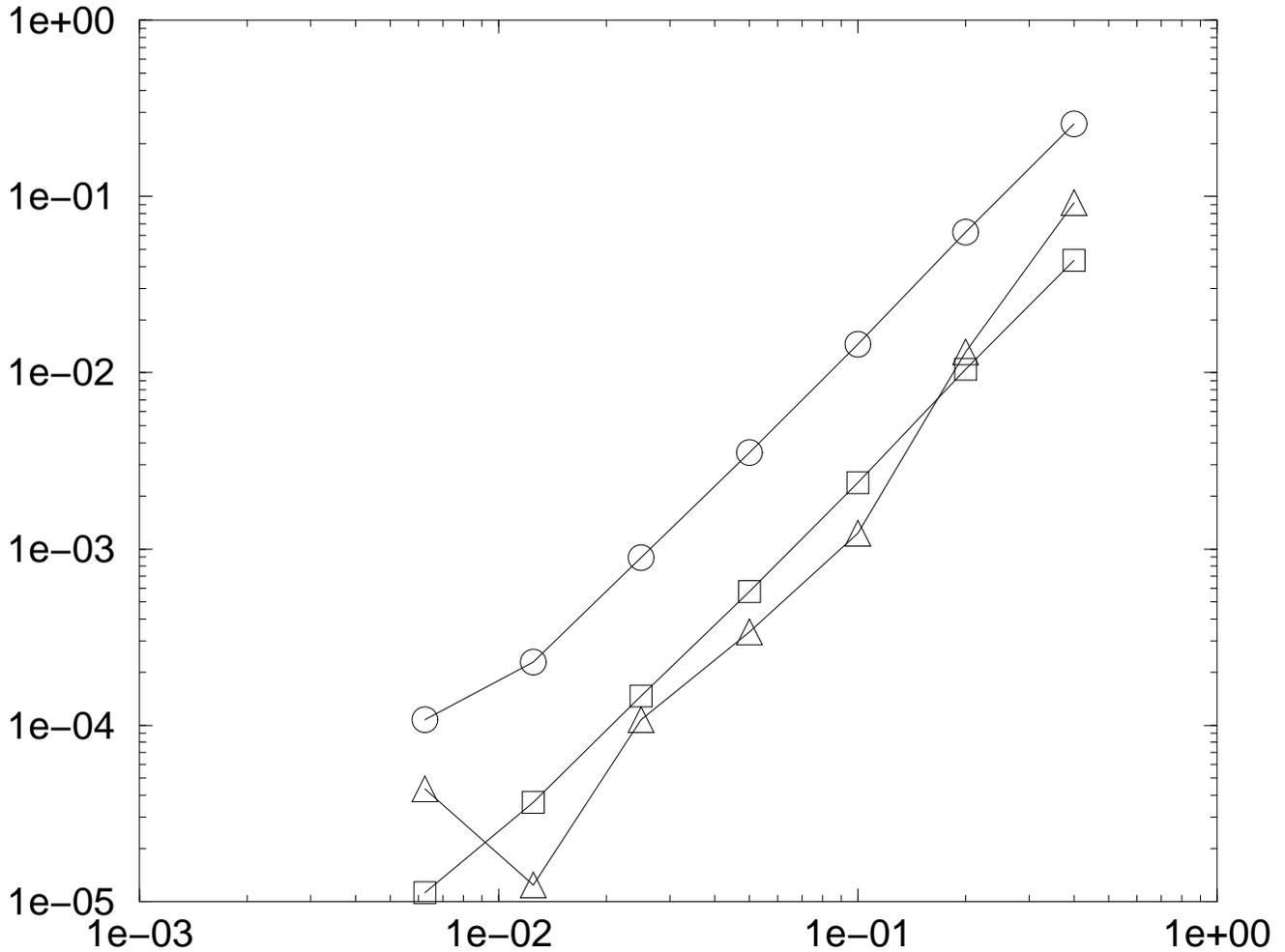}}
\caption{Quadratic convergence of  $Z_{*0}$ with $\Delta\zeta$.
As a measure of the error at $\Delta\zeta$, we compare with
$\Delta\zeta/2$.  Circles denote the maximal error, over all
gridpoints and Fourier components, squares the root mean square error,
and triangles the error in $\Delta$.  $N=64$ and
$\zeta_{\text{left}}=-6.4$. The production value $\Delta\zeta=(1/80)$
corresponds to the second point from the left.}
\label{convergence_m_64}
\end{figure}

Fig. \ref{convergence_m_64} demonstrates quadratic convergence with grid
spacing in $\zeta$, as expected from centered differencing of the
$\zeta$-derivatives. This convergence breaks down at very small values of
$\Delta\zeta$, probably because grid points get very close to the regular
singular point $\zeta=0$.

Convergence with $N$ is rapid: The difference between results for
$N=64$ and $N=128$ is already of order $10^{-6}$.  $N=64$
is surprisingly small, given that it means only 16 odd Fourier
components each to represent $\Pi_+(\tau)$ and $\Pi_-(\tau)$ and 16
even components for $g(\tau)$ and 15 for $\xi_0(\tau)$. (The component
$\cos(4\pi/\Delta)$ of $\xi_0(\tau)$ is taken to be zero to fix the
translation invariance in $\tau$ of the equations for $Z_{*0}$.)

For the production run we have chosen $\zeta_{\text{left}}=-6.4$, $\Delta
\zeta = (1/80)$ (that is, 513 grid points) and $N=128$. 
The solution $Z_{*0}(\zeta,\tau)$ has an estimated maximal error of
$\pm 2.3 \cdot 10^{-4}$ and root-mean-square error of $\pm 3.6 \cdot
10^{-5}$, in the region $-6.4\le
\zeta \le 0$. We obtain $\Delta = 0.73784 \pm 0.00002$. All three error
estimates are dominated by the error from finite differencing in
$\zeta$, with the estimated error from expanding around
$\zeta=-\infty$ somewhat smaller, and the error from using a finite
number of Fourier components in $\tau$ much smaller.

%%%%%%%%%%%%%%%%%%%%%%%%%%%%%%%%%%%%%%%%%%%%%%%%%%%%%%%%%%%%%%%%%%%%%%%%

\section{Linear perturbations and critical exponent}

\subsection{The eigenvalue problem}

In this section we construct the one linear perturbation of the
critical solution that grows with decreasing spacetime scale, as $\tau
\to -\infty$, with the purpose of calculating the critical exponent
for the black hole mass in critical collapse.

The evolution equations for a linear perturbation $\delta Z$ of the
background critical solution $Z_*$ are of the general form
\begin{equation}
\delta Z_{,\zeta} = A \ \delta Z_{,\tau} + (B + e^\tau C 
+ e^{2\tau} D) \ \delta Z.
\end{equation}
The constraints are of the same general form, but with the left-hand
side equal to zero, and the following considerations apply equally to
them as well. The perturbation equations differ from those for the
scalar field model through the explicit appearance of $e^\tau$ in the
equations, and the fact that the coefficients $A$, $B$, $C$ and $D$
are not periodic in $\tau$ because the background solution
$Z_*$ is not. Like $Z_*$, the coefficients $A$, $B$, $C$ and $D$
derived from it admit an expansion of the form
\begin{equation}
\label{A_n}
A=\sum_{n=0}^\infty e^{n\tau} A_n(\zeta,\tau),
\end{equation}
where the $A_n$ are periodic. In this expansion, the leading terms $A_0$,
$B_0$, etc. depend only on the leading term $Z_{*0}$ of the background
expansion.

As for the scalar field model, we make the ansatz \cite{G2}
\begin{equation}
\label{imag1}
\delta Z(\zeta,\tau) = \sum_{i=1}^{\infty} C_i \
e^{\lambda_i\tau}\, \delta_i Z(\zeta,\tau),
\end{equation}
where the $C_i$ are free coefficients, and the $\lambda_i$ are a discrete
set of complex numbers, which are determined as eigenvalues of a new,
linear boundary value problem. 
Clearly the $\delta_i Z$ obey the equations
\begin{equation}
\label{delta_i}
\delta_i Z_{,\zeta} = A \ \delta_i Z_{,\tau} + (B + \lambda_i A + e^\tau C 
+ e^{2\tau} D) \ \delta_i Z.
\end{equation}
In the massless scalar field model, the $\delta_i Z$ could be assumed to be
periodic in $\tau$. In the presence of a scale, this is no longer possible,
and we have to expand each $\delta_i Z$ once more as \cite{GunMar}
\begin{equation}
\label{imag2}
\delta_i Z(\zeta,\tau) = \sum_{n=0}^\infty e^{n\tau} \ \delta_{in}
Z(\zeta,\tau),
\end{equation}
where only the individual coefficients $\delta_{in} Z$ are periodic. This
expansion is exactly analogous to (\ref{scaling}). The $\delta_{in} Z$
obey a coupled set of equations which can be derived from (\ref{delta_i})
in a straightforward bookkeeping exercise, after inserting the expansion
(\ref{A_n}).  These equations are complemented by regularity conditions at
$\zeta=-\infty$ and $\zeta=0$. The equations for the $\delta_{i0}Z$ are
simply
\begin{equation}
\label{delta_i0}
\delta_{i0} Z_{,\zeta} = A_0 \ \delta_{i0} Z_{,\tau} + (B_0 + \lambda_i A_0)
\ \delta_{i0} Z.
\end{equation}
This equation, together with the boundary conditions, already
determines the spectrum $\{\lambda_i\}$. The other $\delta_{in}Z$ obey
inhomogeneous equations and can be determined recursively, but here we
are interested only in the spectrum. This also means that we only need
the leading term $Z_{*0}$ of the background expansion.

Writing down the field equations (\ref{delta_i0}) for the
$\delta_{i0}Z$ is straightforward. As we have seen, one simply
linearizes eqns.  (\ref{background_Pi}-\ref{background_S}) for
$Z_{*0}$, and then makes the replacement $\delta Z_{*0,\tau} \to
\delta_{i0} Z_{,\tau} + \lambda_i \,
\delta_{i0} Z_{,\tau}$, which follows from the definition (\ref{imag1}).
Writing $a$ for $a_{*0}$, etc., and $\delta a$ for $\delta_{i0} a$,
etc., to keep the notation simple, we obtain
\begin{eqnarray}
\label{deltaPi}
\nonumber 
\delta \Pi_{\pm,\zeta} = && 
\left[1 \mp (1+\xi_0') e^{\zeta + \xi_0}\right]^{-1} 
\Bigl\{\mp e^{\zeta+\xi_0} \left[\Pi_{\pm,\tau} 
\ \delta g + g(\delta \Pi_{\pm,\tau} + \lambda_i \ \delta \Pi_\pm) \right]
+ (1-a^2+4a^2S^2) \ \delta\Pi_\pm
\\ 
&& +2a(4S^2-1)\Pi_\pm\ \delta a + 8 a^2 S \Pi_\pm \ \delta S
\mp 2 a^2 \ \delta S \mp 4 a S \ \delta a 
\pm (1+\xi_0') e^{\zeta+\xi_0} \Pi_{\pm,\zeta} \ \delta g \Bigr\},
\\
\delta g_{,\zeta} = && (1-a^2+4a^2S^2) \ \delta g + 2a(4S^2-1) g \ \delta a
+ 8 a^2 S g \ \delta S, \\
\nonumber
0 = && \delta a_{,\tau} + \left[\lambda_i 
+ {1\over2} e^{-(\zeta+\xi_0)} g^{-1}
(\Pi_+^2 - \Pi_-^2) - {1\over2} (1+\xi_0') (1 - 3 a^2 + 12a^2 S^2 + \Pi_+^2
+ \Pi_-^2) \right] \delta a \\
\nonumber
&& + \Biggl\{ e^{-(\zeta+\xi_0)} \left[-{1\over2} 
(\Pi_+^2 - \Pi_-^2) g^{-2} a
\ \delta g
+ g^{-1} a (\Pi_+ \ \delta \Pi_+ - \Pi_- \ \delta \Pi_-) \right] \\
\label{daconstr}
&& - (1+\xi_0') \left[4 a^3 S \ \delta S 
+ a (\Pi_+ \ \delta \Pi_+ + \Pi_- \ \delta \Pi_-) \right] \Biggr\}, \\
\label{dSconstr}
0 = && \delta S_{,\tau} + (\lambda_i + 1 + \xi_0') \ \delta S
+ {1\over2}\left\{ 
e^{-(\zeta+\xi_0)} \left[g^{-2} (\Pi_+ + \Pi_-) \ \delta g - g^{-1} 
(\delta \Pi_+ + \delta \Pi_-) \right] 
+ (1+\xi_0') (\delta \Pi_+ - \delta \Pi_-) \right\}.
\end{eqnarray}
Similarly, we obtain the expansion around $\zeta=-\infty$ of the
$\delta_{i0}Z$ by linearizing eqns. (\ref{left_first}-\ref{left_last})
and then making the same replacement, at each order in $e^\zeta$. The
nonvanishing expansion coefficients to $O(e^{3\zeta})$ are:
\begin{eqnarray}
\delta S_{(1)}(\tau) = && {\rm free}, \\ 
\delta \Phi_{(1)}(\tau) = && - 2 \ \delta S_{(1)}, \\
\delta a_{(2)}(\tau) = && 4 S_{(1)} \ \delta S_{(1)}, \\
\delta \Pi_{(2)}(\tau) = && e^{\xi_0} \left[\delta S_{(1)}' 
+ (\lambda-1-\xi_0') \ \delta S_{(1)} \right], \\
\delta S_{(3)}(\tau) = && {1\over 10} \left\{ e^{\xi_0} \left[\delta
\Pi_{(2)}' + (\lambda -2 - 2\xi_0') \ \delta \Pi_{(2)}\delta 
\right] + 24 S_{(1)}^2 \ \delta S_{(1)}  \right\}, \\
\delta \Phi_{(3)}(\tau) = && - 4 \ \delta S_{(3)}.
\end{eqnarray}

As the linearized regularity condition at $\zeta=0$ we impose the vanishing
of the numerator of (\ref{deltaPi}, upper sign). There is no
linearized equivalent of the coordinate condition (\ref{coordcond}), as we
have fixed the coordinate system already when the background was
calculated. (In other words, $\zeta=0$ is null only on the background
spacetime, not on the perturbed spacetime.) The one boundary condition at
$\zeta=0$ can be solved recursively in terms of two free periodic functions
$\delta g_{(0)}(\tau)$ and $\delta \Pi_{-(0)}(\tau)$, from
\begin{eqnarray}
\label{ds0}
&& \delta S_{(0)}' + (\lambda_i + 1 + \xi_0')\ \delta S_{(0)} + (1+\xi_0')
\left[-\delta \Pi_{-(0)} + {1\over2} g^{-1} \left(\Pi_{+(0)} 
+ \Pi_{-(0)}\right) 
\ \delta g_{(0)}\right] = 0,
\\
\nonumber
&&  \delta a_{(0)}' + \left[\lambda_i + (1+\xi_0')\left(-\Pi_{-(0)}^2 
- {1\over2}+{3\over2} a_{(0)}^2
- 6 a_{(0)}^2 S_{(0)}^2\right)\right]\ \delta a_{(0)} \\
&& 
\label{da0}
- (1+\xi_0') \left[{1\over 2} g_{(0)}^{-1} a_{(0)} 
\left(\Pi_{+(0)}^2 - \Pi_{-(0)}^2\right) \ \delta g_{(0)} 
+ 4a_{(0)}^3 S_{(0)}\ \delta S_{(0)}
+ 2 a_{(0)} \Pi_{-(0)}\ \delta \Pi_{-(0)} \right] = 0, \\
\nonumber
&& \delta\Pi_{+(0)}' + \left[\lambda_i - (1+\xi_0')
\left(1-a_{(0)}^2+4a_{(0)}^2S_{(0)}^2\right) \right]\ 
\delta\Pi_{+(0)} + g_{(0)}^{-1}\left[\Pi_{+(0)}' 
- (1+\xi_0') \Pi_{+(1)}\right]\ \delta g_{(0)} \\
&& 
\label{dpp0}
+ (1+\xi_0') \left\{
2 a_{(0)} \left[\left(1-4S_{(0)}^2\right) \Pi_{+(0)} 
+ 2 S_{(0)}\right]\ \delta a_{(0)}
+ 2 a_{(0)}^2 \left[1-4S_{(0)}\Pi_{+(0)}\right]\ \delta S_{(0)}\right\} = 0.
\end{eqnarray}
The suffix $(0)$ denotes the leading term in an expression in powers
of $\zeta$ around $\zeta=0$. We still need to calculate the background
term $\left(\Pi_{+,\zeta}\right)_{(0)} = \Pi_{+(1)}$ in
eqn. (\ref{dpp0}). To do this, we expand eqns. (\ref{background_Pi},
lower sign), (\ref{6}) and (\ref{4}) to $O(\zeta)$, evaluate the
resulting algebraic expressions
\begin{eqnarray}
g_{(1)} = && C_{(0)} g_{(0)}, \quad \text{where} \quad 
C_{(0)} \equiv 1 - a_{(0)}^2 + 4 a_{(0)}^2 S_{(0)}^2,\\
\Pi_{-(1)} = && {1\over2} \left[ (1+\xi_0')^{-1} \Pi_{-(0)}'
+ C _0 \Pi_{-0} + 2 a_{(0)}^2 S_{(0)}\right], \\
S_{(1)} = && - S_{(0)} - {1\over2}\left(\Pi_{+(0)}-\Pi_{-(0)}\right), \\
a_{(1)} = && {a_{(0)}\over 2} \left(C_{(0)}
+\Pi_{+(0)}^2+\Pi_{-(0)}^2\right),
\end{eqnarray}
(alternatively, we could have obtained $S_{(1)}$ and $a_{(1)}$ from
expanding the constraints (\ref{daconstr}) and (\ref{dSconstr})), and
finally solve the linear ODE
\begin{eqnarray}
\Pi_{+(1)}' && - (1+\xi_0') (1 + 2C_{(0)}) \ \Pi_{+(1)} 
+ (1 + C_{(0)}) \Pi_{+(0)}' \\
&& 
+ (1+\xi_0') \left\{
2 a_{(0)} \left[\left(1-4S_{(0)}^2\right) \Pi_{+(0)} 
+ 2 S_{(0)}\right]\ a_{(1)}
+ 2 a_{(0)}^2 \left[1-4S_{(0)}\Pi_{+(0)}\right]\ S_{(1)}\right\}= 0
\end{eqnarray}
for $\Pi_{+(1)}$.

Linear perturbations which have the same $\tau$-symmetry as the background
$Z_{*0}$ ($S$ and $\Pi_\pm$ odd frequencies, $a$ and $g$ even frequencies)
decouple from those with the opposite symmetry. We call them even and odd
perturbations respectively, and can treat them separately in the numerical
calculation of the spectrum $\{\lambda_i\}$. 

\subsection{Numerical construction}

Our numerical method is the same as in \cite{G2}. We evolve a basis of
all linear perturbations compatible with the constraints at either one
of the boundaries to a matching point, and look for zeros of the
determinant of that basis as a function of $\lambda$. A zero indicates
the existence of a perturbation consistent with both sets of boundary
conditions for that value of $\lambda$. We have implemented this
algorithm for both real and complex $\lambda$. We have checked our
results, for real $\lambda$ and even perturbations, with a relaxation
algorithm, which is partially independent numerically, and in which
$\lambda$ figures as an additional variable, which is balanced by
fixing the perturbations as an additional boundary condition.  The
determinant in question is in fact a holomorphic function of $\lambda$
(because the field equations are real), and this can be used to find
its zeros and poles efficiently.

We expect certain zeros and poles in the $\lambda$-plane from the following
considerations. $Z_{*0}$ is scale-invariant, and therefore invariant under
the infinitesimal transformation
\begin{equation}
\label{gauge1}
Z_{*0}(r,t) \to Z_{*0}[(1+\epsilon)r, (1+\epsilon)t] \simeq
Z_{*0}(\zeta,\tau+\epsilon) \simeq
Z_{*0}(\zeta,\tau) + \epsilon Z_{*0,\tau}.
\end{equation}
This corresponds to a gauge linear perturbation mode with $\lambda_i=0$ and 
$\delta_i Z = Z_{*0,\tau}$. $Z_{*0}$ is also invariant under time
translation, 
\begin{equation}
\label{gauge2}
Z_{*0}(r,t) \to Z_{*0}(r,t+\epsilon) \simeq Z_{*0}(\zeta,\tau) 
+ \epsilon \ e^{-\tau} \left[
(1+\xi_0') Z_{*0,\zeta} -Z_{*0,\tau}\right],
\end{equation}
corresponding to a gauge mode with $\lambda_i=-1$. Both gauge modes are
even according to our classification.

The ODEs, eqns. (\ref{ds0}, \ref{da0}, \ref{dpp0}), are all of the form
$f'+gf+h=0$, where $f$ stands for $\delta S_{(0)}$, $\delta a_{(0)}$
and $\delta \Pi_{+(0)}$ respectively. In all three equations $g$
depends only on the background solution and is even, while $h$ is
linear in the perturbations, and has the same $\tau$-symmetry as $f$.
It can be shown \cite{G2} that this type of equation has no solution
when the average value (in $\tau$) of the coefficient $g$ vanishes. As
$g$ in each case is of the form $\lambda+(\text{background fields})$,
this corresponds to a simple pole in the $\lambda$-plane.
These poles are not just due to the breakdown of a particular
numerical method but indicate that for these values of $\lambda$ no
perturbations exist which obey the boundary condition at $\zeta=0$.
The poles arise only when the inhomogeneous term $h$, and in
consequence the unknown $f$, have a nonvanishing average, that is when
they are even.

Calculating the average value of $g$ for each of the three equations, we
find that they vanish for $\lambda=-1$, $\lambda=-1-A$ and $\lambda=-A$
respectively, where $A$ is the average value of $2(1+\xi_0')\Pi_{-(0)}^2$,
with numerical value $A\simeq 0.1726$. (We have used the fact that 
\begin{equation}
(\ln
a_{(0)})'=(1+\xi_0')\left[\Pi_{-(0)}^2+{1\over2}
+\left(2S_{(0)}^2-{1\over2}\right)a_{(0)}^2\right]
\end{equation}
has vanishing average value, as it is the derivative of a periodic
function, to simplify the averages.)  In summary, for even
perturbations we expect zeros at $\lambda=0$ and $\lambda=-1$ (gauge
modes), one more zero on the negative real line (the unstable mode)
and a pole at $\lambda\simeq -1.17$.  For odd perturbations we expect
poles at $\lambda=-1$ and $\lambda \simeq -0.17$.

The numerical calculation of the perturbation determinant as a
function of $\lambda$ largely confirms the predictions: For even
perturbations, on the negative real line we find a zero at $\lambda_1
\simeq -5.0$, corresponding to the expected physical unstable mode,
with a critical exponent of $-1/\lambda_1 \simeq 0.2$, as found in
collapse simulations.  We also find the expected zero at $\lambda_2 =
0$. We have verified that the corresponding $\delta_i Z
\propto Z_{*0,\tau}$ to high precision. 
We find the expected pole at $\lambda\simeq-1.17$, but accompanied by
a zero very close by.  For odd perturbations, on the negative real
line we find the expected pole at $\lambda\simeq-0.17$.

At $\lambda=-1$, for both even and odd perturbations, we do
not find the expected zero and pole respectively because of a
numerical problem which is discussed in the appendix. It does not
affect our calculation of the perturbation determinant for values of
$\lambda$ not close to $-1$.  The unstable mode at $\lambda\simeq-5.0$
and gauge mode at $\lambda=0$ are clear enough, and we can use their
convergence properties to obtain an estimate of the numerical error.

Table 1 gives the values of $\lambda$ for the unstable mode
$\lambda_1$ and the scale change gauge mode $\lambda_2$ as a
function of the step size $\Delta\zeta$. The deviation of the
numerical value of $\lambda_2$ from zero serves as one estimate of
numerical error. It is larger than the other estimate, from the
convergence of $\lambda_1$, and we therefore adopt it as our
definitive error estimate for $\lambda_1$. We obtain $\lambda_1 =
-5.091 \pm 0.017$, from which we obtain for the critical exponent
$\gamma = -1/\lambda_1 = 0.1964 \pm 0.0007$.

%%%%%%%%%%%%%%%%%%%%%%%%%%%%%%%%%%%%%%%%%%%%%%%%%%%%%%%%%%%%%%%%%%%%%%%%

\section{Conclusions}

We have obtained the asymptotic form of the type II critical solution of
Einstein-Yang-Mills collapse, its echoing period, and the critical
exponent for the black hole mass, in a calculation similar to the one
we made for the massless scalar field
\cite{G1,G2}. The major new feature is the presence of the length scale
in the Einstein-Yang-Mills field equations. In consequence, the
critical solution and its linear perturbations are no longer
self-similar, but become so only asymptotically on spacetime scales
much smaller than the length scale of the field equations (on the
order of the Bartnik-McKinnon \cite{BM} mass). Here we have only
calculated the leading term in the asymptotic expansions for the
critical solution and its perturbations, but this is sufficient to
calculate both the echoing period $\Delta$ and critical exponent
$\gamma$ exactly. We find $\Delta = 0.73784 \pm 0.00002$ and $\gamma =
0.1964\pm 0.0007$, while Choptuik, Chmaj and Bizo\'n \cite{EYM} find
$\Delta\simeq 0.74$ and $\gamma \simeq 0.20$ in collapse simulations.
Data files of the background and unstable mode from the production run
are available through the WWW address
http://www.aei-potsdam.mpg.de/$\sim$gundlach.

In the formalism we have developed here to deal with the presence of a
length scale in the equations, the leading perturbation term, $\sum
C_i e^{\lambda_i\tau} \delta_{i0} Z$, obeys field equations which are
the linearized version of the equations for the leading background
term, $Z_{*0}$. Both sets of equations consistently describe a new
physical system which is scale-invariant, and which is obtained from
the original, scale-dependent, model in the limit where all fields
vary on spacetime scales much smaller than the intrinsic scale of the
field equations. In the language of renormalisation group theory,
these equations are the short-scale fixed point of a renormalisation
group transformation acting on the original field equations. In the
case of a massless or self-interacting scalar field this fixed point
is the massless scalar field \cite{Koike2}. For scalar
electrodynamics, the fixed point is the massless complex scalar
without electromagnetism
\cite{GunMar}.
In both cases the field equations at the fixed point can naturally be
associated with the Lagrangian of a simpler physical system, and the
renormalisation group acts naturally on the dimensionful coupling
constant. While the latter is still true for the spherically symmetric
Einstein-Yang-Mills system, the equations
(\ref{background_Pi}-\ref{background_S}) are not the spherical
reduction of some set of covariant field equations. The reduction to
spherical symmetry does not commute with the action of the
renormalisation group.

We believe that in the present paper we have developed the most
general formalism that will be be required to deal with critical
collapse restricted to spherical symmetry, in allowing for
self-similarity and the presence of a length scale. The generalization
to more than one scale is trivial: the various scales can be written
as a single scale times dimensionless numbers. The general formalism
has already given rise to a bit of new physics: the calculation of
critical exponents not only for the black hole mass but also its
charge in critical collapse of scalar electrodynamics
\cite{GunMar}.

The example of Einstein-Yang-Mills collapse shows that one does not
need scale-invariance of the field equations to have type II critical
phenomena with the famous relation $M\sim(p-p_*)^\gamma$. Rather they
can be found in some region of initial data space for any system. For
astrophysical matter, these initial data are simply not realized in
astrophysical collapse.

Most remaining questions in critical phenomena go beyond the
restriction to spherical symmetry. Do the spherical critical solutions
found so far act also as critical solutions for generic,
non-spherical, initial data? What is the angular momentum of the black
hole formed from data with angular momentum in the limit where the
black hole mass is fine-tuned to zero? Are there qualitatively new
phenomena away from spherical symmetry?

%%%%%%%%%%%%%%%%%%%%%%%%%%%%%%%%%%%%%%%%%%%%%%%%%%%%%%%%%%%%%%%%%%%%%%%%

\acknowledgments

I would like to thank Piotr Bizo\'n and Matt Choptuik for helpful
discussions, and Matt Choptuik for making his data from collapse
simulations available. This work was supported by a scholarship of the
Ministry of Education and Science (Spain).

%%%%%%%%%%%%%%%%%%%%%%%%%%%%%%%%%%%%%%%%%%%%%%%%%%%%%%%%%%%%%%%%%%%%%%%%

\appendix

\section{Numerical problems at $\lambda\simeq-1$}

Calculating the determinant of even perturbations as a function of
$\lambda$, we do not find the expected zero but a pole at $\lambda =
-1$, with an alternating series of poles and zeros accumulating
towards $-1$ from below. There are no further poles or zeros
immediately above. The positions of the poles are the same in the real
and complex algorithms, but depend on the values of the numerical
parameters $N$ and $\delta\zeta$. A qualitatively similar picture
arises for odd perturbations. These features can be explained as a
numerical artifact as follows.

We have checked explicitly that the gauge mode (\ref{gauge2}) obeys
the constraint (\ref{daconstr}). When we try to reconstruct $\delta a$
of this mode from the constraint, however, the numerical result blows
up at small $\zeta$.  To understand this, consider the equation
$\delta a' + g\ \delta a + h = 0$, with $g$ and $h$ defined by
eqn. (\ref{daconstr}). The Fourier algorithm that we use to solve this
for $\delta a$ at each $\zeta$ needs to divide the average of $h$ by
the average of $g$.  As $\zeta \to -
\infty$, the average of $g$ over $\tau$ as a function of $\lambda$ and
$\zeta$ is $\lambda + 1 + O(e^{2\zeta})$, where the last term is
positive. As $\lambda\to-1$ from below, this goes through zero at some
small value of $\zeta$. In the exact perturbation mode (\ref{gauge2}),
the average value of $h$ vanishes at the same rate with $\zeta$ as
that of $g$, but with small numerical errors this cancellation fails,
and small numerical errors are magnified. In the calculation of the
perturbation determinant this results in the observed, essentially
random behavior for $\lambda
\lesssim -1$. For $\lambda>-1$ the problem does not arise, as then the
average value of $g$ does not vanish for any $\zeta$.

We have not found a simple way of fixing this problem, as our
algorithm relies in an essential way on reconstructing $S$ and $a$ and
$\delta S $ and $\delta a$ from the constraints at each $\zeta$. It
does not affect numerical results however unless where the average
values of both the coefficients $g$ and $h$ are very small, that is
for $\lambda
\lesssim -1$. (If only the average value of $g$ is small,
the resulting blowup in the perturbations is physical, as in the other
poles we have discussed.) Calculating the perturbation determinant is
not goal in itself, but only a means of finding the spectrum of linear
perturbations. With the present method we can say with confidence that
there is a zero at $\lambda\simeq -5.0$, and no other zeros for
negative real $\lambda$, apart from the two gauge modes. We could in
principle be missing a zero (physical growing mode) at
$\lambda\lesssim -1$, where the code is unreliable, and have to rely
on evidence from collapse simulations that there is only one unstable
mode.

%%%%%%%%%%%%%%%%%%%%%%%%%%%%%%%%%%%%%%%%%%%%%%%%%%%%%%%%%%%%%%%%%%%%%%%%

%%%%%%%%%%%%%%%%%%%%%%%%%%%%%%%%%%%%%%%%%%%%%%%%%%%%%%%%%%%%%%%%%%%%%%%%%

%%%%%%%%%%%%%%%%%%%%%%%%%%%%%%%%%%%%%%%%%%%%%%%%%%%%%%%%%%%%%%%%%%%%%%%%%%%

%%%%%%%%%%%%%%%%%%%%%%%%%%%%%%%%%%%%%%%%%%%%%%%%%%%%%%%%%%%%%%%%%%%%%%%%%%%

\begin{table}
\caption{Convergence of $\lambda$ with step size in $\zeta$.
$\lambda_1$ is the Lyapunov exponent of the
one growing mode. Its negative inverse is the critical exponent
$\gamma$. $\lambda_2$ is the exponent of the scale change
($\tau$-translation) gauge mode. It
must be zero and serves as a check on the numerical error. Note that the
numerical grid (and number of steps) is the same for the background as for
the perturbations in each case. The range of $\zeta$ is $-6.4\le\zeta\le0$
in each case.}
\label{table2}
\begin{tabular}{ccc}
Number of steps & $\lambda_1$ & $\lambda_2$ \\ 
\tableline
32 & -5.1318584162589 & 0.13720816860828 \\
64 &  -5.0828194924873 & -1.4932102080912E-02 \\
128 & -5.0891816495598 & -1.1519697001693E-02 \\
256 & -5.0910562847918 & 4.7061190684163E-03 \\
512 & -5.0913625725286 & 1.6758611037677E-02 \\
\end{tabular}
\end{table}

%%%%%%%%%%%%%%%%%%%%%%%%%%%%%%%%%%%%%%%%%%%%%%%%%%%%%%%%%%%%%%%%%%%%%%%%%

\end{document}